\documentclass[useAMS,usenatbib,printer,psfig,epsf]{aa}
\usepackage{natbib,graphicx}

\begin{document}

\title{Time-series Spectroscopy and Photometry of the Pulsating
Subdwarf B Star PG~1219+534 (KY~UMa)\thanks{Based on observations from the
Nordic Optical Telescope, Kitt Peak National Observatory, and MDM Observatory.}}\titlerunning{Time-series Spectroscopy of PG~1219+534}

\author{M. D. Reed \inst{1}\thanks{Visiting Astronomer, Kitt Peak National Observatory, National Optical Astronomy Observatory, which is operated by the Association of Universities for Research in Astronomy (AURA) under cooperative agreement with the National Science Foundation.} \and J. R. Eggen\inst{1} \and S. L. Harms\inst{1} \and J. H. Telting\inst{2}
\and R. H. \O stensen\inst{3} \and S. J. O'Toole\inst{4}
\and D. M. Terndrup\inst{5} \and A.-Y. Zhou\inst{1,6} \and R. L. 
Kienenberger\inst{1} \and U. Heber\inst{7}}
\offprints{M. D. Reed}

\institute{Department of Physics, Astronomy, \& Materials Science, Missouri State University, 901 S. National, Springfield, MO 65897 U.S.A. \email{mreed@sdbv.missouristate.edu, joe.astro@gmail.com, trillianlala@gmail.com, aiyingzhou@gmail.com, r.vanwey@gmail.com}
\and Nordic Optical Telescope, Apartado 474, 38700 Santa Cruz de La Palma,
Spain \email{jht@not.iac.es}
\and Institute of Astronomy, Catholic University of Leuven, 
Celestijnenlaan 200B, 3001 Leuven, Belgium \email{roy@ster.kuleuven.ac.be}
\and Anglo-Australian Observatory, PO Box 296, Epping NSW 1710, Australia
\email{otoole@aao.gov.au}
\and Department of Astronomy, The Ohio State University, Columbus, OH 43210
U.S.A. \email{terndrup@astronomy.ohio-state.edu}
\and National Astronomical Observatories of the Chinese Academy of
Sciences, Beijing 100012,P.R. China 
\and Dr. Remeis-Sternwarte Bamberg, Universit\"{a}t Erlangen-N\"{u}rnberg, Sternwartstr. 7, 96049 Bamberg, Germany \email{Ulrich.Heber@sternwarte.uni-erlangen.de}}

\date{Received \today / Accepted \today}

\abstract{}{
We present observations and analysis 
of time-series
spectroscopy and photometry of the pulsating subdwarf
B  star PG~1219+534 (KY~UMa). Subdwarf B stars are blue horizontal
branch stars which have shed most of their hydrogen envelopes. Pulsating
subdwarf B stars allow a probe into this interesting phase of evolution.}
{Low resolution spectra were obtained at the Nordic
Optical Telescope and Kitt Peak National Observatory, and photometric
observations were obtained at MDM and Baker observatories in 2006. 
We extracted radial velocity and equivalent width variations from
several Balmer and He~I lines in individual spectra. The pulsation 
frequencies were separated via
phase binning to detect line-profile variations in Balmer and helium
lines, which were subsequently matched to atmospheric models to infer
effective temperature and gravity changes throughout the pulsation cycle.}
{ From the
photometry we recovered the four previously observed frequencies and detected
a new fifth frequency. 
From the spectra we directly measured radial velocity and 
equivalent width variations for the four main frequencies and from 
atmospheric models we successfully inferred temperature and gravity changes
for these four frequencies.
We compared amplitude ratios and phase differences of these quantities
and searched for outliers which could be identified as high-degree modes.
These are the first such measurements for a ``normal'' amplitude pulsating      subdwarf B star, indicating that spectroscopic studies can benefit the majority
of pulsating subdwarf B stars.
}{}

\maketitle

\section{Introduction}
Subdwarf B (sdB) stars are the field counterparts to extended horizontal
branch stars observed in globular clusters. Their masses are 
about 0.5~M$_{\odot}$
with thin ($<$$10^{-2}$M$_{\odot}$), inert hydrogen envelopes and temperatures
from $22\,000$ to $40\,000$~K 
(Heber et al. 1984; Saffer et al. 1994), making them very blue.
Their origin remains a mystery as mass loss on the first-ascent giant
branch must remove all but
$< 10^{-2}$M$_{\odot}$ of the hydrogen envelope, yet produce a core of the 
same mass (to within a few percent) every time 
\citep{cruz,han02,han03}.
Pulsating sdB stars come in two varieties. The short period (90 to 600
seconds) pulsators were discovered in 1997 \citep{kill97} and were 
named EC~14026 stars after that prototype. Their official designation is
V361~Hya stars,  the General Catalog of Variables Stars lists them as RPHS
\citep{kaz}, but they are typically referred to as
sdBV stars. Their usual amplitudes are near 1\% of their mean brightness
and stars have been detected with as few as one 
 (HS~1824+5745 \citep{reed06a})
 or as many as 55 (PG~1605+072 \citep{kill99})
pulsation frequencies \citep{reed07a}. 
Long period (45 minutes to 2 hours) pulsators were discovered
in 2003 \citep{grn03,mdr2004_2} and are officially designated as V1093~Her
stars but commonly named PG~1716 stars after that prototype, PG~1716+426. 
Their amplitudes are typically $<$0.1\% and they tend to be multimode
pulsators. 
For this work, our interest is the sdBV
class of pulsators as their periods are short, so many pulsation
cycles can be observed during each run from a single site. The short
periods also produce spectroscopic variations (radial velocity, equivalent
width, effective temperature and gravity) at an observable level for 
intermediate resolution spectrographs and 4~m class telescopes.

Pulsating sdB stars potentially allow the opportunity to discern
their interior structure using asteroseismology to obtain estimates
of total mass, luminosity, envelope mass,
radiative levitation, gravitational
diffusion, and helium fusion cross sections. To do so,
the pulsation modes must first
be identified with their associated spherical harmonics.
In 1999, \citet{simon0,simon2,simon3} began applying
time-series spectroscopic techniques to attempt
mode identification in sdBV stars. Whereas photometry measures 
brightness changes largely caused by temperature variations, 
spectroscopy can reveal information regarding the pulsation
velocities and separate the temperature and gravity components. An
excellent example of these techniques was shown in \citet{to04} where they
determined that the pulsational degree
 of the highest-amplitude frequency of PG~1325+101 (QQ~Vir) was
consistent with $\ell\,=\,0$.
Other sdBV stars studied using time-series spectroscopy include
PB~8783 (EO~Cet) and KPD~2109+4401 (V2203~Cyg) \citep{jp00}, PG~1605+072
(V338~Ser) \citep{simon0,simon2,simon3,simon,woolf},
and Balloon~090100001 \citep{to06} using low-resolution spectroscopy
on intermediate-sized telescopes.
PG 1219+534, PG 1605+072, and PG 1613+426 were also observed 
using FUSE \citep{kua}. Only for the high-amplitude pulsators (PG~1605+072,
PG~1325+101, and Balloon~090100001) were pulsations detected. However,
such pulsators provide their own problems in that the high-amplitude
pulsations affect the other frequencies and therefore require considerable
prewhitening (variation removal); e.g. \citet{to06,till}.
Unfortunately, because of insufficient signal, temporal resolution, or
run length, none of these studies have provided conclusive mode 
identifications which would be useful for constraining models.

This paper reports our time-series spectroscopic and photometric
observations of PG~1219+534 (also KY~UMa but hereafter PG~1219). 
PG~1219 was discovered to
be a pulsator by \citet{koen99b} who detected four independent frequencies.
The spacings of the frequencies
are such that the entire pulsation spectrum can be resolved in just a 
few hours, yet the frequency density is too large for the pulsations to be
of the same modal degree $\ell$. A photometry-based mode identification
has been published which attributes the four frequencies to three 
differing modal degrees \citep{charp05a} which can be used as a guide for our
mode identifications. While the pulsation amplitudes are 
low (2 -- 8~mma in photometry), they should not interfere with each other,
eliminating the need to prewhiten some frequencies to reveal others. 
This makes PG~1219 an excellent target for
mode identification studies.
Our photometric observations of PG~1219 began in 2003 as part of our effort
to resolve the pulsation spectra (Fourier transform; FT) of sdBV
stars \citep{me2,reed06a,reed07a,reed07b,zhou}; \citet{harms} reported
the results of 2003 -- 2005 photometric
observations of PG~1219. During that time, the
pulsation frequencies remained consistent with those of
\citet{koen99b} and occasionally a new, low-amplitude frequency would
appear for one or two nights.

 In \S 2 we describe our simultaneous
spectroscopic and photometric
observations during 2006 and examine the frequency content in
\S 3. In \S 4 we discuss the results of our observations 
and in \S 5 we provide
conclusions from our work. Further detailed modeling
to understand our observational results in terms of asteroseismology
will be presented in a separate paper.

\section{Observations}
\subsection{Photometry}
We obtained photometry at MDM and Baker observatories to
support our spectroscopic observations, the details of which are
provided in Table~\ref{tab01}. Baker observatory (BO) is equipped with a 0.4~m 
telescope and a Roper Scientific RS1340b CCD photometer.
At MDM we used the 1.3~m McGraw Hill telescope
with an Apogee Instruments U47 CCD. For all of
our CCD measurements, we binned the CCDs $2\times 2$ pixels providing a 
dead time of 1 second and all used a red
cut-off filter (BG38 or BG40), so the effective bandpass covers the $B$
and $V$ filters and is essentially that of a blue-sensitive photomultiplier
tube. Accurate timing was accomplished via Network
Time Protocol (NTP) connections to at least one stratum 1 and two stratum
2 servers. NTP timings were accurate to better than 0.02~s.

\begin{table*}
\caption{2006 Photometric observations of PG~1219 simultaneous or near to
the spectroscopic observations. \label{tab01}}
\centering
\begin{tabular}{lccccc}\\ \hline \hline
Run & Date & Start Time & Length  & Number of  & Observat.\\
Name & (UT) & (UT) & (hours) & images &  \\ \hline
mdm040806    & 8 Apr.  & 3:10:30     &   9.3 & 3900   &  MDM \\
mdm040906    & 9 Apr. & 3:21:30     &   4.4 & 1916   &  MDM \\
mdm041006    & 10 Apr. & 2:24:52     &   2.8     & 521     &  MDM \\
mdm041106    & 11 Apr. & 4:23:30     &   3.5 & 1132    &  MDM \\
bak041806    & 18 Apr. &  5:09:30    &   4.9 & 1583   &  BO \\
bak041906    & 19 Apr. & 2:23:30     &   7.7 & 2056  &  BO \\
bak042006    & 20 Apr. & 2:20:00     &   6.6 & 1815  &  BO \\
bak042206    & 22 Apr. & 2:25:00     &   7.4 & 2030  &  BO \\ \hline
\end{tabular}
\end{table*}

\begin{figure}
\includegraphics[width=\columnwidth]{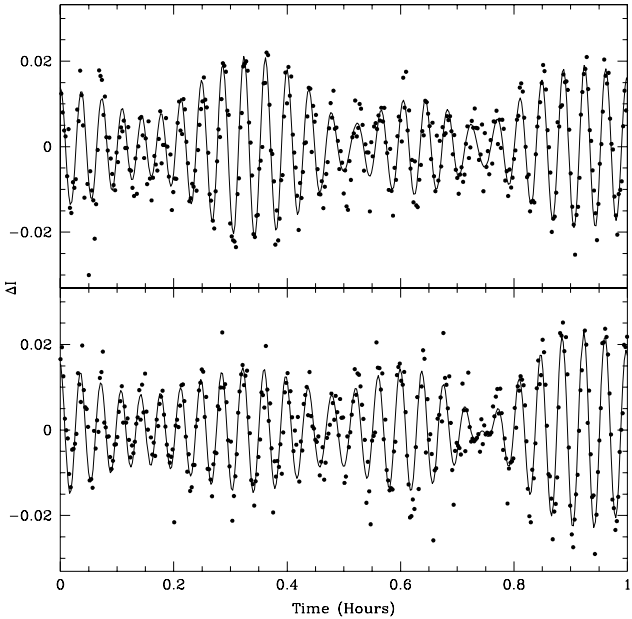}
\caption{Photometric lightcurves for consecutive 1 hour segments of PG~1219.
Solid line is a four-frequency fit to the points.}
\label{fig01}
\end{figure}

Our photometric data were reduced using 
standard {\sc iraf} packages for image reduction, including 
bias subtraction,
dark current and flat field corrections.
Intensities were extracted using {\sc iraf} aperture photometry with extinction
and cloud corrections using the normalized intensities of several field
stars. As sdB stars are substantially hotter than typical field
stars, differential light curves
are not flat due to differential atmospheric and colour extinctions.
A low-order polynomial was fit to remove these trends from the data on a
night-by-night basis. Finally, the lightcurves are normalized by their
average flux and centered around zero so the reported differential
intensities are $\Delta {\rm I}=\left( {\rm I}/\langle{\rm I}\rangle\right)-1$.
Amplitudes are given as milli-modulation amplitudes (mma) with an
amplitude of 10~mma corresponding to 1.0\% or 9.2~millimagnitudes.
A portion of
data from MDM is shown in Fig.~\ref{fig01}.
There is obvious beating that occurs on multiple time scales which
indicates the multiperiodic nature of the pulsations. 

\subsection{Spectroscopy}
During 7 nights in April 2006 we obtained nearly 5200 
time-resolved low-resolution spectra of PG~1219, with the specifics
of each run provided in Table~\ref{tab02}. We obtained 2085
spectra at the Kitt Peak National Observatory's (KPNO) 4~m Mayall
telescope using the RC spectrograph in long-slit mode. Our KPNO
instrumental setup was: grating KPC-007 with the T2KB CCD which 
sampled approximately 3000-5000~\AA , a FWHM spectral resolution of
about 4~\AA , and a dispersion of 1.44~\AA /pixel. We subframed
the CCD along the slit, reading out a section $2080\times 37$ including
a 41 column overscan region. Integration times were 10~s with rather
long dead-times of 25~s for a total cycle time of $\sim 35$~s. 
Binning the CCD provided only negligible improvement, so we did not.
We obtained helium-neon-argon arc-line spectra every 80-150 spectra
aperiodically and observed the flux-standard star Feige~66 three times 
per night. Halogen lamp flat fields were obtained nightly for removing 
pixel-to-pixel variations on the CCD.

 \begin{table}
\caption{Spectroscopic Observations of PG~1219.}
\label{tab02}
\centering
\begin{tabular}{lccccc} \hline\hline
Date & Start Time & Length  & Number of & Observat.\\
2006 & (UT) & (hours) & spectra & \\ \hline
 8 Apr. & 3:23 & 8.9  & 810 & KPNO \\
 9 Apr. & 2:47 & 9.5  & 850 & KPNO \\
11 Apr. & 7:34 & 4.7  & 425 & KPNO \\
17 Apr. & 20:43 & 8.3  & 757 & NOT \\
18 Apr. & 20:32 & 8.9  & 760 & NOT \\
19 Apr. & 20:29 & 8.7  & 826 & NOT \\
20 Apr. & 20:31 & 8.0  & 756 & NOT \\ \hline
\end{tabular}
\end{table}

We obtained 3099 spectra at the Nordic Optical Telescope (NOT) using
{\sc alfosc} in long-slit mode. Our NOT setup used 
grism \#16 and CCD \#8, which sampled approximately 3500-5050~\AA ,
a resolution of about 3~\AA~ FWHM at a dispersion of
0.77~\AA /pixel. The grism and slit were set up such that they were aligned
with the rows of the CCD to shorten the dead times. Using 25~s integrations,
we achieved a cycle time of $\sim$31~s, so while the NOT integrated longer
than KPNO, the cycle time was similar. A few re-acquisitions were done per
night to correct the slit angle with the parallactic angle. Thorium-argon
and helium arc-line spectra were obtained every 60-80 spectra. Halogen
lamp images were obtained for flat-fielding.

\begin{figure}
\includegraphics[angle=-90,width=\columnwidth]{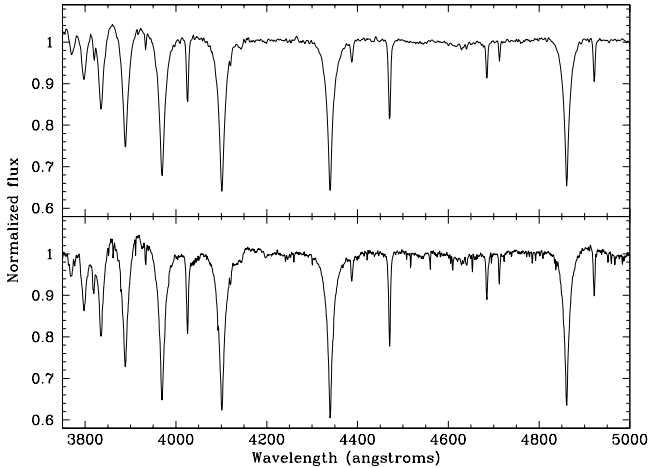}
\caption{Average of all normalized KPNO (top) and NOT (bottom) spectra.}
\label{fig02}
\end{figure}

The spectra were bias and dark-current corrected using overscan regions
and flat field corrected using standard tasks within {\sc iraf}. Two
bad columns in the NOT data were corrected by linear interpolation of 
pixels in adjacent columns. As the spectra contains low signal-to-noise
(S/N) regions in the UV, and the KPNO CCD had a sharp sensitivity decline
redward of ~5030~\AA , the spectra were trimmed to 3710-5130~\AA $\,$
for the KPNO data and 3480-5010~\AA $\,$
for the NOT data. One-dimensional spectra were 
optimally extracted after subtracting a fit to the sky background for each 
detector column. Wavelength calibration was done using the HeNeAr and the
ThAr calibration spectra for the KPNO and NOT data, respectively,
interpolating the wavelength solution between the nearest before and after
calibration spectra. The average S/N of the spectra was 31 for KPNO and
25 for the NOT data, so while the KPNO telescope was much larger than the
NOT, the longer NOT integrations, combined with a more efficient grating
allowed similar S/N but higher  spectral resolution in the NOT data. 
The spectra were normalized to the continuum using
the {\sc iraf} task {\sc continuum} which used a fourteenth order
Legendre polynomial. The mean normalized KPNO and NOT spectra are shown
in Fig.~\ref{fig02} on the same wavelength scale.
The 10 and 25~s integration times lead to phase smearing that will
reduce amplitudes of variation by 0.85 and 5.2\%, respectively, for a
140~s pulsation period. Additionally, our measurements are for disk-averaged
observations and intrinsic amplitudes may be higher locally within the
unresolved surface.

\section{The frequency content of the pulsations}

\begin{table*}
\caption{Frequencies and amplitudes detected in our annual photometric
 data sets.
All frequencies are given in $\mu$Hz and amplitudes in mma with least-squares
errors on the last digits in parentheses.}
\label{freqlist}
\centering
\begin{tabular}{lcccccccc} \hline\hline
ID & \multicolumn{2}{c}{2003} & \multicolumn{2}{c}{2004} &\multicolumn{2}{c}{2005} &\multicolumn{2}{c}{2006} \\ 
 & Freq. & Amp. & Freq. & Amp. & Freq. & Amp. & Freq. & Amp. \\ \hline
$f1$ & 6721.484(10) & 2.8(1) & 6721.529(55) & 3.9(2) & 6721.481(158) & 2.2(4) & 6721.591(4) & 2.9(1) \\
$f2$ & 6961.363(6) & 5.3(1) & 6961.355(33) & 6.6(2) & 6961.202(52) & 6.8(4) & 6961.489(2) & 6.4(1) \\
$f3$ & 7490.021(9) & 3.6(1) & 7489.727(38) & 5.7(2) & 7492.850(52) & 6.8(4) & 7490.833(2) & 4.5(1) \\
$f4$ & 7807.738(8) & 4.0(1) & 7807.771(29) & 7.4(2) & 7807.733(36) & 9.7(4) & 7807.857(2) & 6.1(1)\\
$f5$ & & & 7398.56(30) & 0.7(2) \\
$f6$ & & & & & & & 7744.6(45) & 2.1(4)\\
$f7$ & & & 8168.83(17) & 1.2(2) \\ \hline
\end{tabular}
\end{table*}

\begin{figure}
\includegraphics[width=\columnwidth]{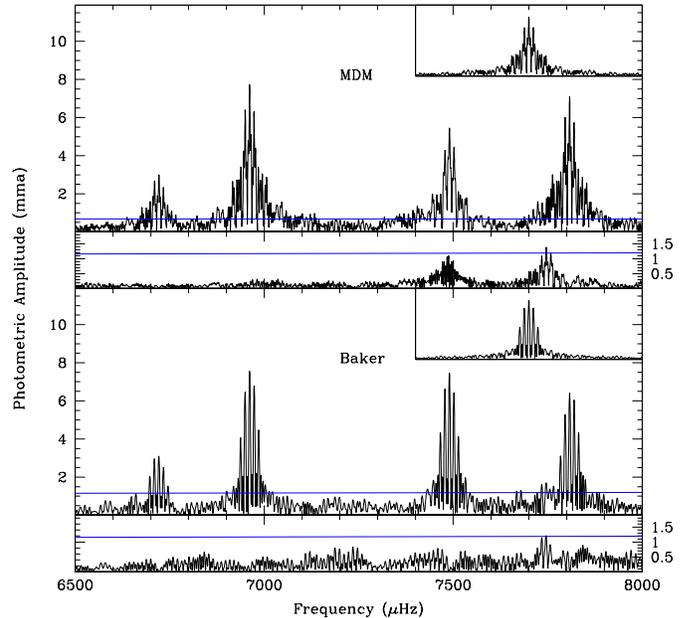}
\caption{Pulsation spectra of photometric data simultaneous with
the spectroscopic observations. Smaller panels have the four main frequencies
removed and insets are the spectral windows plotted on the same frequency
scale. The (blue)
lines indicate the $4\sigma$ detection limit.}
\label{phot4spec}
\end{figure}

Our selection of  PG~1219 as a target is partially due to 
its year-to-year stability and this continued during our 2006 
observations. Table~\ref{freqlist} provides the frequencies and amplitudes
of pulsations detected during our four-year photometric
 program and indicates that two
additional frequencies were observed during 2004, but during single nights
only. However, during our 2006 campaign, a new, low amplitude frequency
persisted through both weeks of observations. These occasionally-observed
frequencies are labeled as $f5$ through $f7$ of Table~\ref{freqlist}.
Pulsation spectra of photometry which overlaps our spectroscopic data
are shown in Fig.~\ref{phot4spec}.  Insets show the spectral
window (FT of a single sine-wave sampled at the same times as the data) and
smaller panels show the residuals after prewhitening by the four main
frequencies. The solid horizontal (blue) line is the $4\sigma$ detection
calculated as four times the average value of the FT for regions
outside of areas of pulsation \citep{breg1}.

Photometric amplitudes of $f1$ through $f4$ show 20-40\% variation over the
course of four years, but during our spectroscopic observations, the
amplitudes remain essentially constant. The
only significant change is $f3$, for which the amplitude increases by 14\%
in the week between the MDM and BO observations.
From our photometric monitoring of PG~1219 we can conclude that there are
four consistently detected frequencies 
and a fifth, low amplitude frequency. 

Since our cycle time and S/N were better for photometry, we could examine
the photometry for nightly variations in phase  (defined as
time of first maximum since JD=2453830.0 divided by the 
pulsation period). Such variations can
cause integrated data to indicate lower amplitudes or have multiple peaks
which are not intrinsic to the star. The phases of the four main frequencies 
were consistent to within the errors of those provided in Table~4, and so
we conclude that no phase variations occurred during the photometric
or spectroscopic runs. However the phases do change between the MDM and
BO observations. The phase of $f3$ changes by 13\% while the others
show marginal changes of 2, 4, and 3\% for $f1$, $f2$, and $f4$, respectively.
 
\subsection{Time-series spectroscopy}
\begin{figure}
\includegraphics[angle=-90,width=\columnwidth]{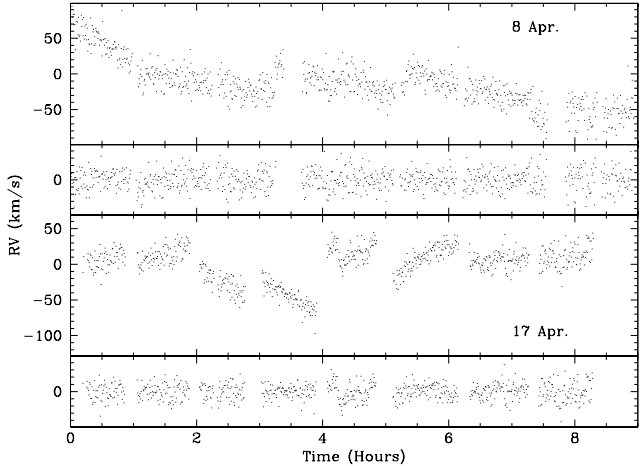}
\caption{Radial velocities obtained from cross-correlation of the time-resolved
spectra. The top two panels show a run from KPNO both before and after
fitting with polynomial segments and the bottom two panels show a run
from the NOT before and after being fitted with polynomial segments. All
panels are plotted to the same relative scale.}
\label{rvLC}
\end{figure}

The procedures we used are similar to those described in \citet{to04} and
summarized as follows:
We calculated the radial velocity (RV) of the time-resolved spectra using the
cross-correlation application {\sc fxcor} in {\sc iraf}. We 
produced nightly template spectra from the mean of individual spectra which
{\sc fxcor} uses to do a Fourier cross-correlation between the template
and individual spectra. We filtered the input to fit line features rather than
large-scale trends in each spectrum, and a Gaussian was fit to
the resulting cross-correlation function (CCF) of a best-fit size between 3 
and 21 velocity bins around the maximum. We fit the CCF for the H$\beta$
through H8 Balmer lines and the 4026 and 4471~\AA\ HeI lines. The radial
velocity shifts are dominated by those of the strongest Balmer lines
in the spectra. Figure~\ref{rvLC} shows radial velocities obtained from
one night each of KPNO and NOT data. The raw velocities, shown in the top
portion for each date, show jumps in velocity due to repositioning
the target on the slit and longer-term instrumental effects. These were
corrected by fitting first or second-order polynomials to the segments
between repositionings and are shown on the bottom portion for each date. All
panels are plotted to the same scale but with varying velocity offsets. As 
these changes were of significantly longer time then the pulsations, there
was no impact on pulsation velocities.

\begin{figure}
\includegraphics[angle=-90,width=\columnwidth]{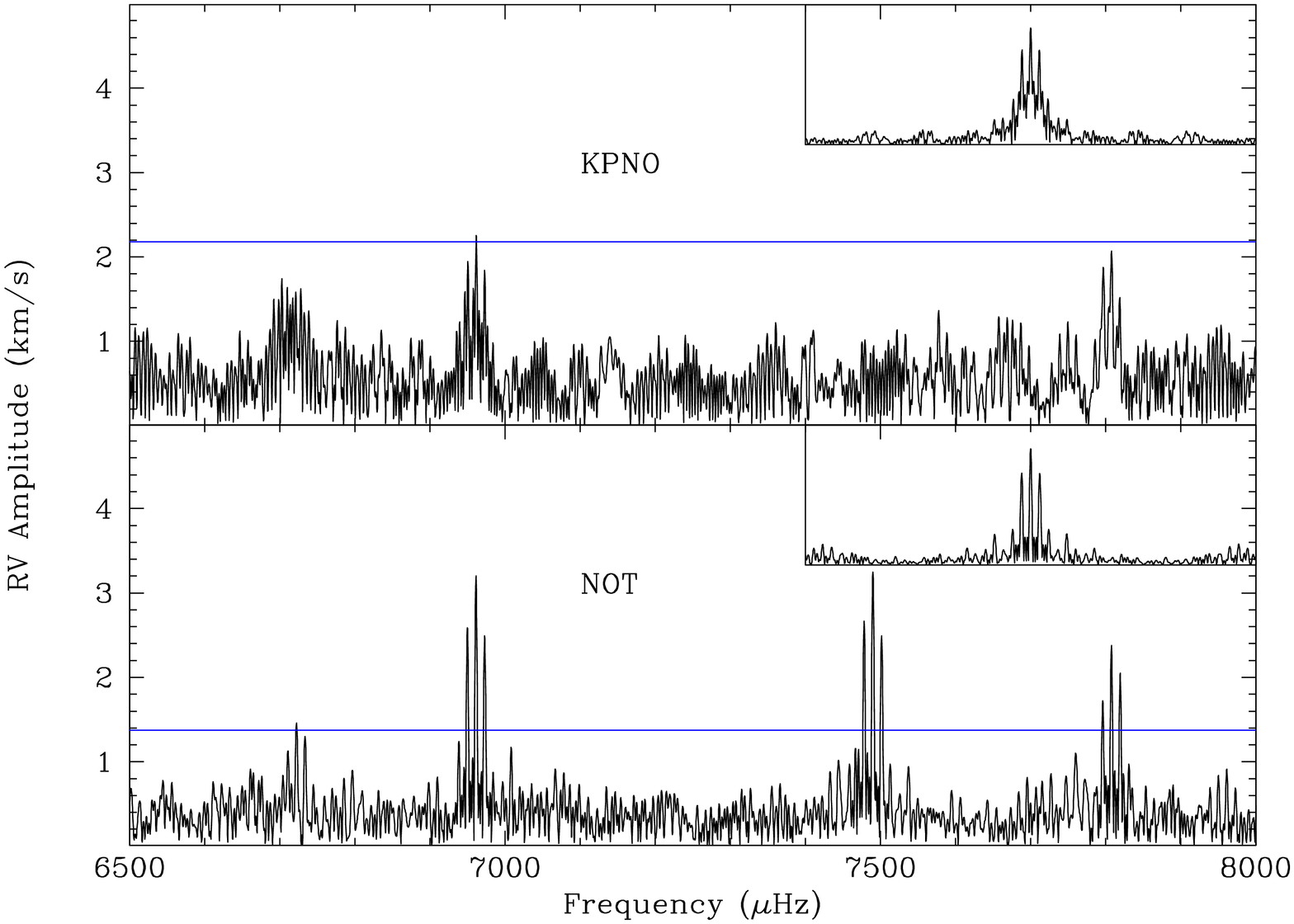}
\caption{Pulsation spectra of radial velocity data. Insets are the 
spectral}
windows and (blue) lines indicate the $4\sigma$ detection limit.
\label{rvFT}
\end{figure}

Figure~\ref{rvFT} shows the temporal spectra of the KPNO and NOT RV data
separately. The solid (blue) line shows the $4\sigma$ detection limit and the
insets show the spectral window. Results from least-squares fitting for the 
amplitudes and phases are provided in Table~\ref{tab04} in \S 4. 
Apparent in the figure is the missing peak where $f3$ should be in the 
KPNO data. While
the $4\sigma$ limit is nearly double that of the NOT data, $f2,$ $f3,$ and
$f4$ have amplitudes in the NOT data which are sufficient to be detected in the
KPNO data. $f2$ and $f4$ are both at least marginally detected, 
yet there is no trace of $f3$ at all. $f1$ shows a noisy 'bump' in the 
appropriate place, but at just over $3\sigma$, it is below the detection limit.

\begin{figure}
\includegraphics[angle=-90,width=\columnwidth]{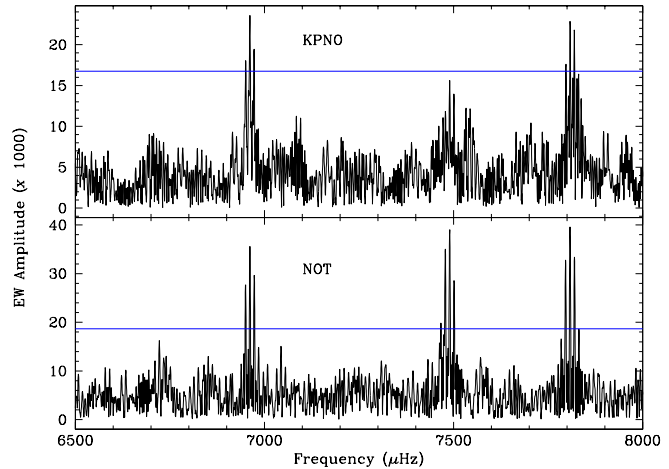}
\caption{Temporal spectra of equivalent width data.}
\label{ewFT}
\end{figure}

We also compute the equivalent widths (EW) of the H$\beta$ 
through H8 Balmer lines.
For our EW measurements,
 we shifted the spectra using the cross-correlation velocities to account
for instrumental and pulsational velocity shifts.
The resulting standard deviation in velocity is 0.07~km/s with an
average CCF fit error of 5.34~km/s.
To reduce the noise in the wings of the profiles, each point in the profile
was weighted with its own depth with respect to the continuum, using the 
following non-standard form, EW$=\sqrt{\Sigma\left( 1-P\left(\lambda\right)
\right)^2}\times \Delta\lambda$. No long-term trends were evident in the
EW data, so no polynomial corrections were made. The results are shown
in Fig.~\ref{ewFT}. $f1$ is below the detection limit for both KPNO and NOT
data, $f2$ and $f4$ are easily recovered in both sets, and $f3$ is clearly
detected in the NOT data, but just below $4\sigma$ in the KPNO data.
As evident in Fig.~\ref{ewFT}, NOT EW values were found to be 
systematically larger than for the
KPNO data. Analysis using various resolutions and line-bin-widths has shown
the amplitudes to be dependent on both the resolution and
the number of pixels used to determine the line widths. Because of these
dependencies it is not feasible to compare the amplitudes between
the two runs without some normalization. 
As the photometric and RV amplitudes of $f4$ are the closest
between the runs, in Table~\ref{tab04} we have normalized the KPNO EW 
amplitudes to that of $f4$ from the NOT run, while we show their
original amplitudes in Fig.~\ref{ewFT}.

\section{Discussion}
\subsection{Effective temperature and gravity variations}
\begin{figure}
\includegraphics[width=\columnwidth]{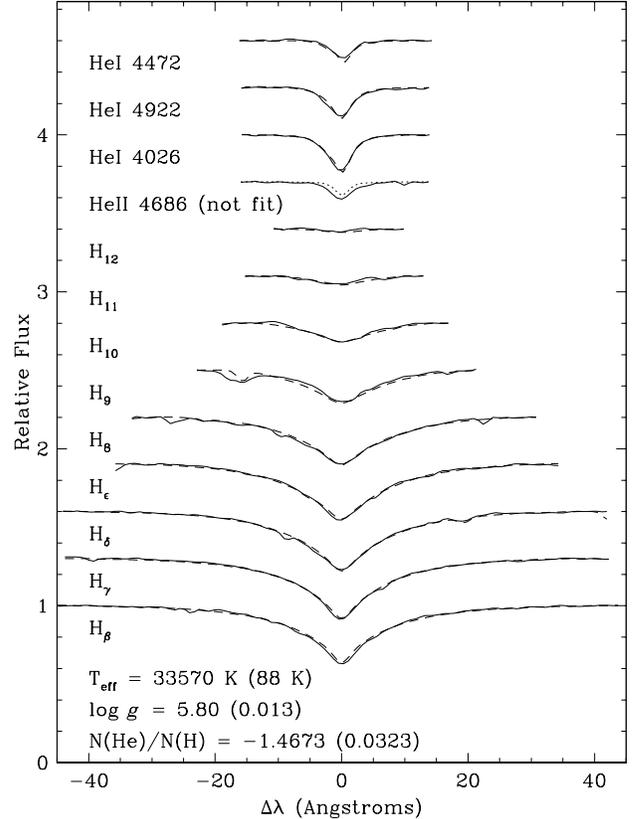}
\caption{Atmospheric models (dashed lines) fitted to the template
spectrum of combined NOT data for several H and He~I lines. The
 He~II line is shown, but not used in the fit. Line identifiers and
resultant atmospheric parameters are
provided at the bottom.}
\label{fitprof}
\end{figure}

To measure changes in $T_{\rm eff}$ and $\log g$ for individual pulsation
frequencies, we folded the spectra into 20 phase bins according to the
pulsation periods. By folding over each pulsation period, we average out
the variations caused by the other three periods, effectively isolating
those of the folded period. The phase-binned spectra have S/N in the
range 125 to 160. We then fit these binned spectra
 with the LTE synthesis model atmospheres
of \citet{heber00}. The fitting is a $\chi^2$ process described by
\citet{berg92} and updated to include helium abundance fits by
\citet{saf94}. These model fits provide values of $T_{\rm eff}$, $\log g$,
 and helium abundance as a function of pulsation phase for
each of the periods. As the helium abundance did not change during
any of the pulsations, we fixed this to the mean value and refit the models,
mildly decreasing the fit errors on $T_{\rm eff}$ and $\log g$. In 
Fig.~\ref{fitprof} we illustrate the fitting procedure by showing the
atmospheric fit to the template spectrum produced by combining all of the 
NOT spectra. The lines used for the fit are shown in the figure and
include several Balmer lines and three He~I lines. A single He~II line
is shown, but was not used in the fit.
The fitted values for $T_{\rm eff}$, $\log g$, and N(He)/N(H) are also
provided on the figure and are nearly within the errors of those previously
published \citep{charp05a,heber00,koen99b,simon6}. 
There are known systematic effects between LTE and NLTE model atmospheres
with and without iron-group diffusion \citep{heber04,simon6} and so we do not
claim any increased accuracy in our values, but merely show the lines
to illustrate the fitting procedure. \citet{simon6} were able to overcome
these inconsistencies using supersolar metal abundances, so we consider
their measurements of $T_{\rm eff}$ and $\log g$ as the most reliable.
In our case we are not really
concerned with the actual values of $T_{\rm eff}$ and $\log g$ but the
changes that occur through a pulsation cycle. In this sense, minor
differences from previously reported values do not contribute to the
fit errors in the variations.

\subsection{Fits of determined quantities}
\begin{figure}
\includegraphics[width=\columnwidth]{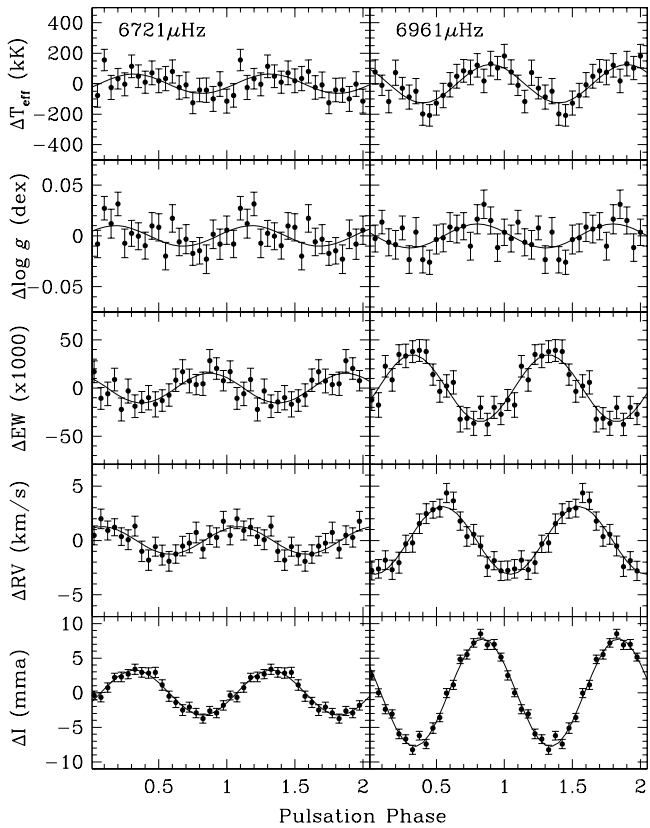}
\caption{Variations in $T_{\rm eff}$, $\log g$, equivalent width, radial 
velocity, and photometry for $f1\,=\,6721$ and $f2\,=\,6961\,\mu$Hz of the 
NOT/BO data, along with non-linear least-squares
fits (solid lines). Phases shown twice for clarity.}
\label{fitP1}
\end{figure}

\begin{figure}
\includegraphics[width=\columnwidth]{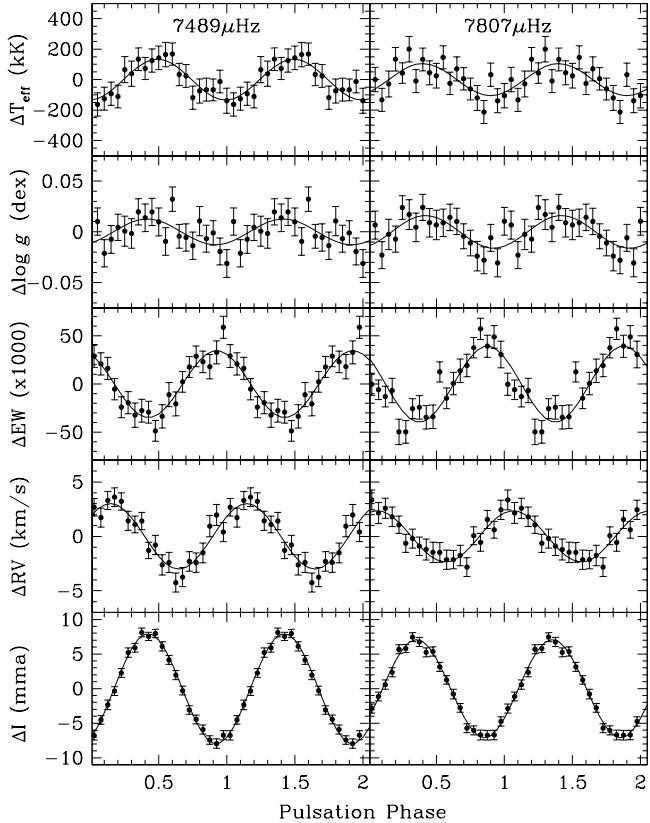}
\caption{Same as Fig.~\ref{fitP1}
for $f3\,=\,7489$ and $f4\,=\,7808\,\mu$Hz of the NOT/BO data.}
\label{fitP3}
\end{figure}

\begin{figure}
\includegraphics[width=\columnwidth]{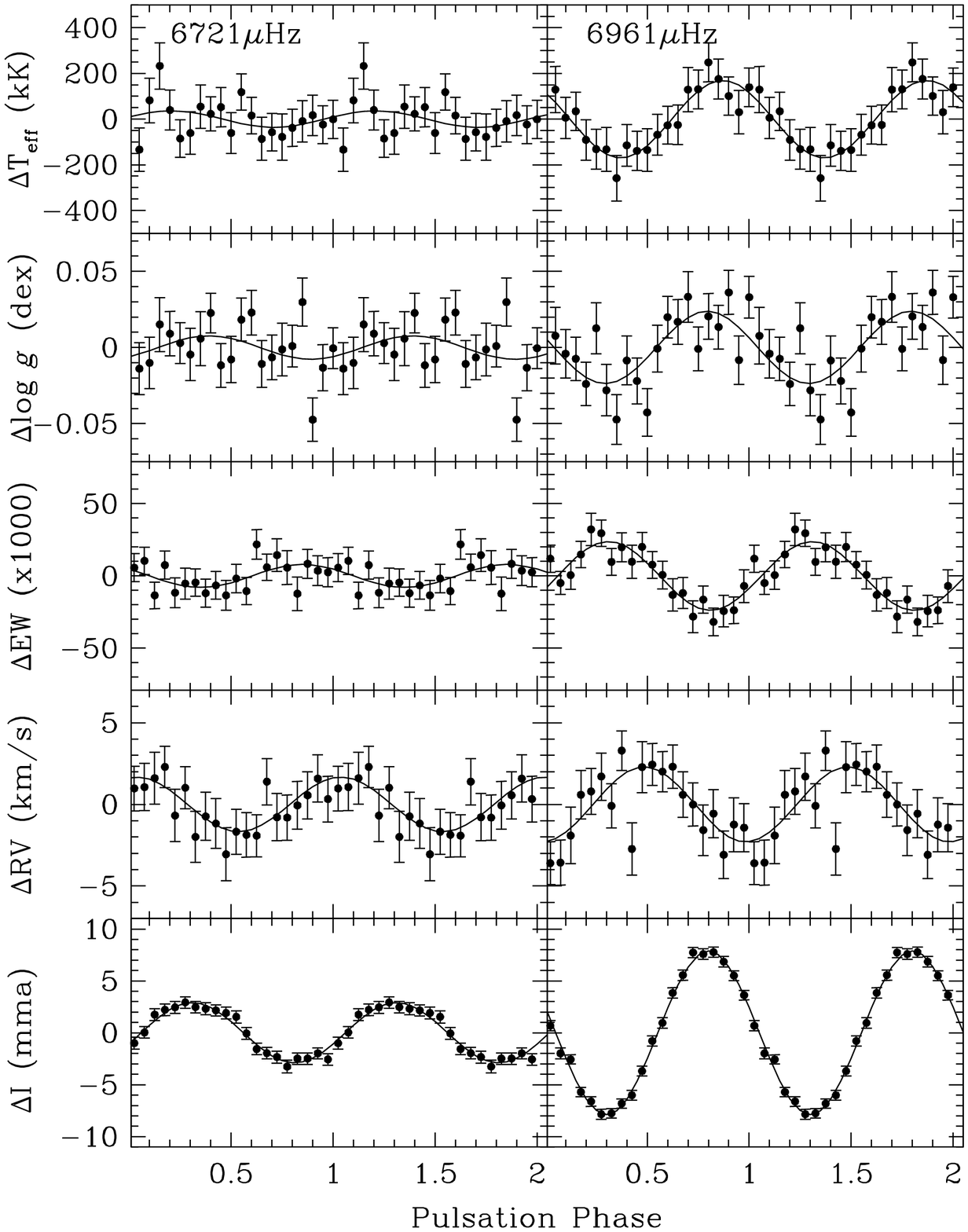}
\caption{Same as Fig.~\ref{fitP1} for KPNO/MDM data.}
\label{fitK1}
\end{figure}

\begin{figure}
\includegraphics[width=\columnwidth]{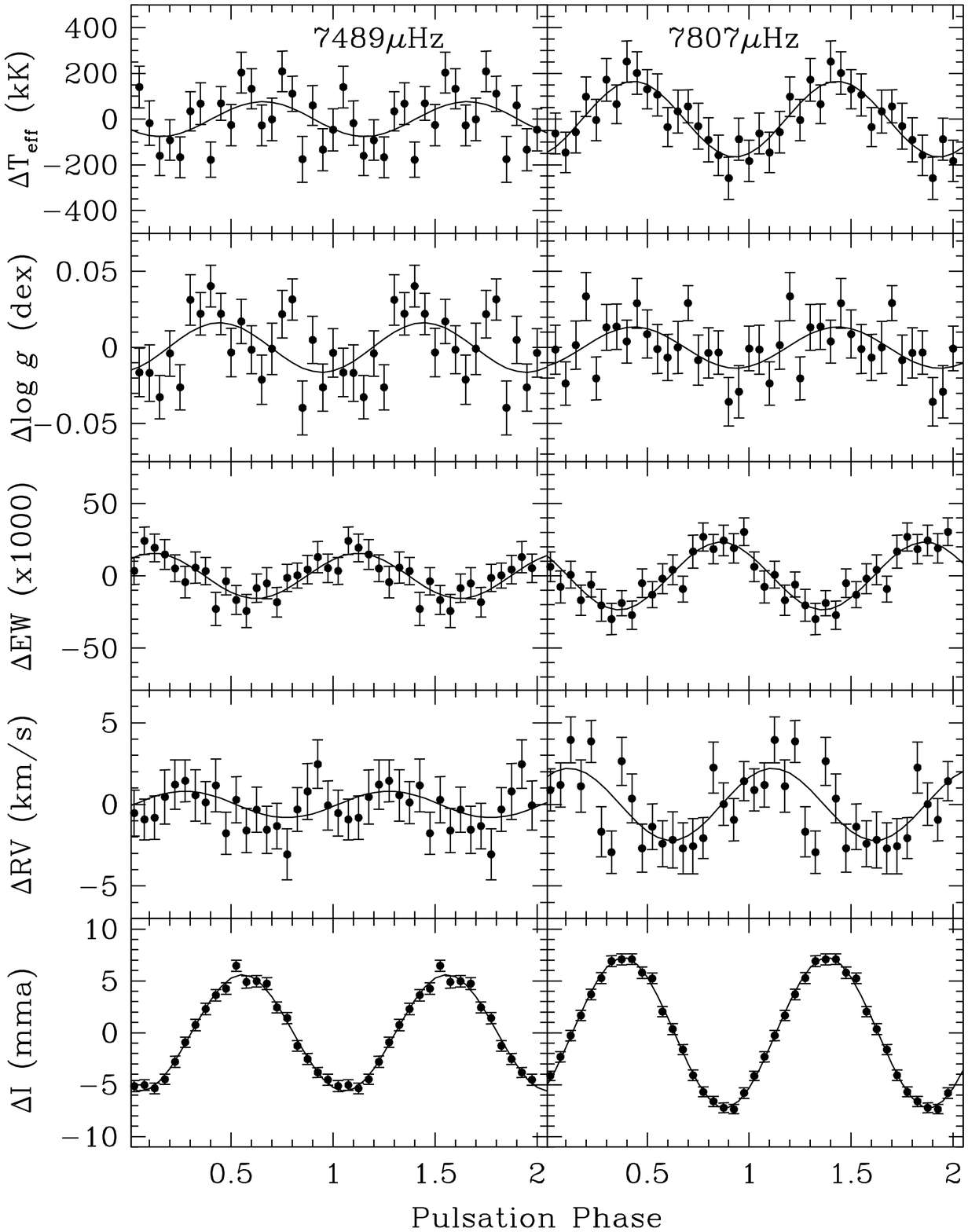}
\caption{Same as Fig.~\ref{fitP3} for KPNO/MDM data.}
\label{fitK3}
\end{figure}

To determine amplitudes and phases of the measurables (or model quantities),
we used non-linear least-squares (NLLS) fits to 
measure the amplitudes and phases
with the pulsation frequencies fixed to the those derived from photometry
(which match those in velocity). For photometry, RV, and EW,
we fit the data themselves with timing corrected to the barycenter of
the solar system. For $T_{\rm eff}$ and $\log g $, these are model fits
to spectra folded over the frequency fixed to the photometric value. Then
the variations in model $T_{\rm eff}$ and $\log g $ were fit using our
NLLS routine. The phase-folded data, along with the
fits are shown in Figs.~\ref{fitP1} through \ref{fitK3} with the quantities
provided in Table~\ref{tab04}. Frequency-folded data were produced from
the photometric and RV data for the plot, but these were not fitted for
the quantities in Table~\ref{tab04} except as noted ($f1$ and $f3$ RV
data for KPNO and $f1$ EW data for KPNO and NOT). 
Table~\ref{tab04} also provides the
$4\sigma$ detection limits from the data.
No such values can be determined from the atmospheric model fits,
though their formal errors have been folded into the NLLS formal errors
shown in the figures and provided in the table.


\begin{table*}
\caption{Results of photometric and spectroscopic variations in
PG~1219. 
Notes: $^a$ indicates quantities were fit directly from the data, though they
were below the $4\sigma$ limit. $^b$ indicates quantities that were fit from
folded data (but not modeled). The KPNO EW amplitudes have been normalized
to that of $f4$ of the NOT data.}
\label{tab04}
\centering
\begin{tabular}{lcccccc} \hline\hline
Measure & Obs &  $f1$ & $f2$ & $f3$ & $f4$ & Limit \\ \hline
\multicolumn{7}{c}{Amplitudes}\\
Photometric  & MDM & 2.9(2)& 7.9(2)& 5.6(2)& 7.2(2)& 0.6 \\
(mma) & BO & 3.3(2) & 7.8(2) & 8.0(2) & 7.1(2) & 1.1  \\
RV & KPNO & 1.6$^a$(5) & 2.3(5) & 0.8$^{a}$(0.5) & 2.2(5) & 2.2  \\
(km$\cdot$s$^{-1}$) & NOT & 1.4(3) & 3.1(3) & 3.0(3) & 2.4(3) & 1.2  \\
EW ($10^{-3}$) & KPNO &13.3$^a$(5.4) & 40.0(5.4) & 26.2$^a$(5.4) & 39.4(5.4) & 28.2  \\
   & NOT & 16.2$^a$(3.8) & 34.7(3.8) & 34.8(3.8) & 39.4(3.8) & 18.2  \\
$T_{\rm eff}$  & KPNO & 46(26) & 175(44) &  58(30) & 159(44)  &  \\
(K) & NOT & 59(26) & 122(34) & 130(34) & 111(35)  &  \\
$\log g$  & KPNO & 12 (7) & 24(7) & 15(8) & 17(6) &  \\
 ($10^{-3}$cm$\cdot$s$^{-2}$)& NOT & 11(5) & 11(5) & 11(5) & 16(5)&  \\
\multicolumn{7}{c}{ Phases }\\
Photometric & MDM & 0.325(10) & 0.798(3) & 0.556(5) & 0.382(4) & \\
 & BO & 0.310(11) & 0.838(5) & 0.423(4) & 0.354(5) &  \\
RV  & KPNO & 0.023$^a$(44) & 0.484(32) & 0.276$^a$(90) & 0.117(33) & \\
        & NOT   & 0.060(33) & 0.545(15) & 0.142(15) & 0.066(19) &   \\
EW & KPNO & 0.837$^a$(65) & 0.310(22) & 0.074(33) & 0.860(22) &  \\
   & NOT & 0.866$^a$(37) & 0.323(17) & 0.926(17) & 0.832(15) & \\
$T_{\rm eff}$ & KPNO & 0.207(117)& 0.820(40) & 0.605(82) & 0.381(41) & \\
              & NOT & 0.316(58) & 0.850(44) & 0.441(41) & 0.349(52) & \\
$\log g$ & KPNO & 0.392(113)& 0.794(51) & 0.444(72) & 0.384(69) & \\
         & NOT & 0.182(74) & 0.802(63) & 0.417(61) & 0.350(52) & \\ \hline
\end{tabular}
\end{table*}

Since we know that the pulsations are phase-stable during our photometric
observations, we can
expect them to be phase-stable for the other measurables, and our results
indicate it to be the case. Even the 13\% change in phase for $f3$ between the 
MDM and BO runs is measured, to within the errors for RV, EW,
$T_{\rm eff}$, and $\log g$ between the KPNO and NOT runs, even though 
some amplitudes show large variations. This indicates that our
phases are measured reliably.

\subsection{Amplitude ratios and phase differencing}
\begin{figure}
\includegraphics[width=\columnwidth]{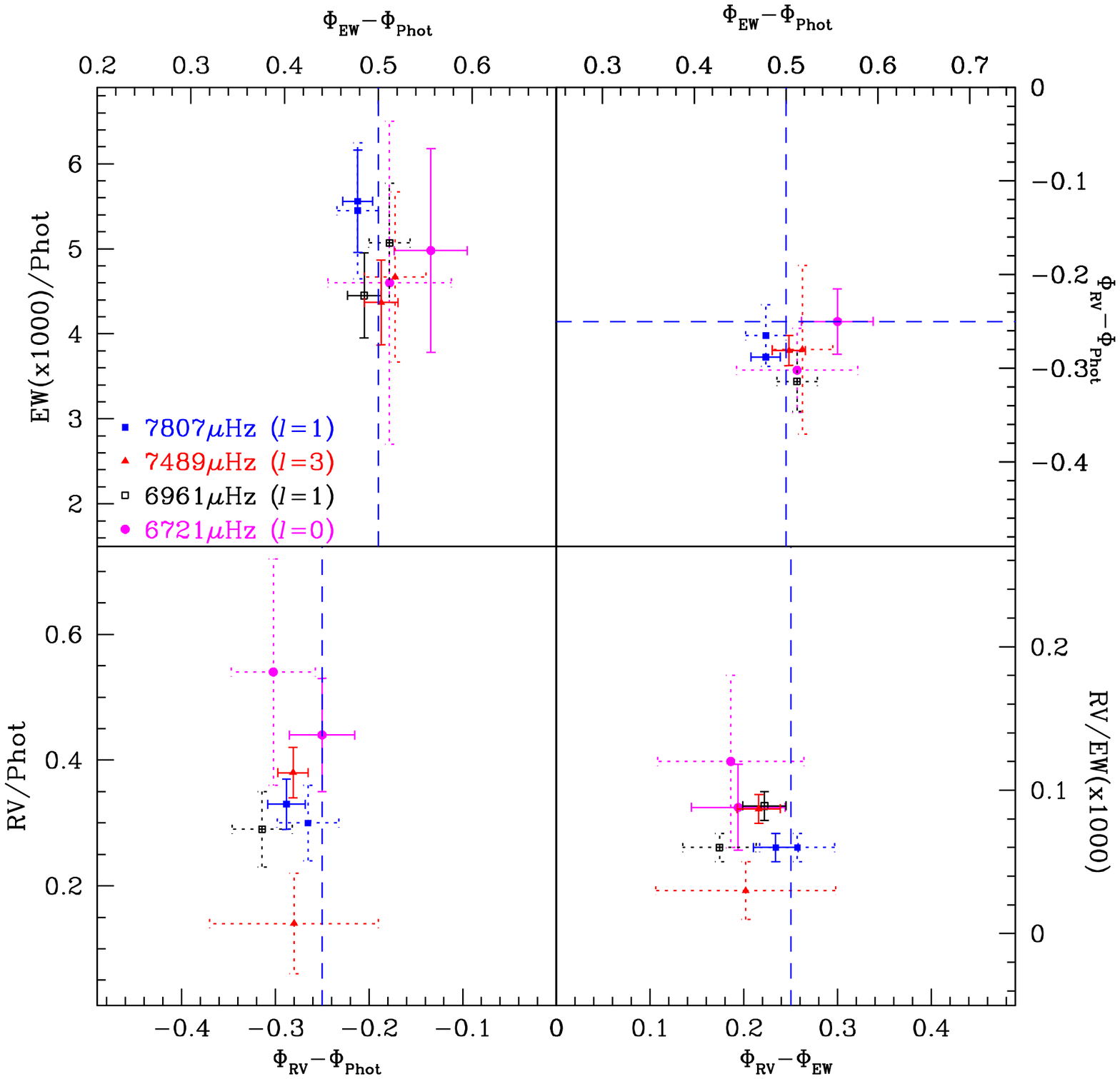}
\caption{Amplitude ratios versus phase differences for fitted observables.
Solid errorbars indicate NOT/BO data while dotted errorbars indicate
KPNO/MDM data. Dashed lines indicate expected phases from
model simulations and mode identifications are those of \citet{charp05a}.}
\label{fig16}
\end{figure}

While we leave any attempts at mode identification to the second paper,
which will properly simulate our results using pulsation models, it is
still interesting to examine what we detected in such a context. It is
expected that pulsations of various modes will behave differently between
photometry, velocity, equivalent width, effective temperature, and gravity
measurements, and 
that such differences
(and similarities) will provide strong constraints on mode identifications. 
There has already been a model for this star
matched to photometric data \citep{charp05a}
which concluded that our four measured frequencies should
be associated with three different pulsation degrees. As such, we should
expect observable differences between our measured quantities.
The mode identifications of \citet{charp05a} are provided in Fig.~\ref{fig16}.

\begin{figure}
\includegraphics[width=\columnwidth]{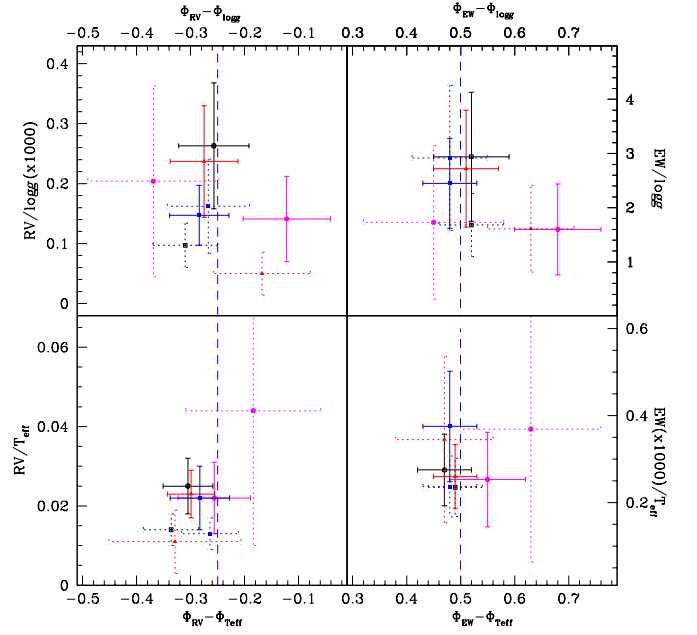}
\caption{Same as Fig.~\ref{fig16} for quantities determined solely from 
spectroscopy.}
\label{fig17}
\end{figure}

\begin{figure}
\includegraphics[width=\columnwidth]{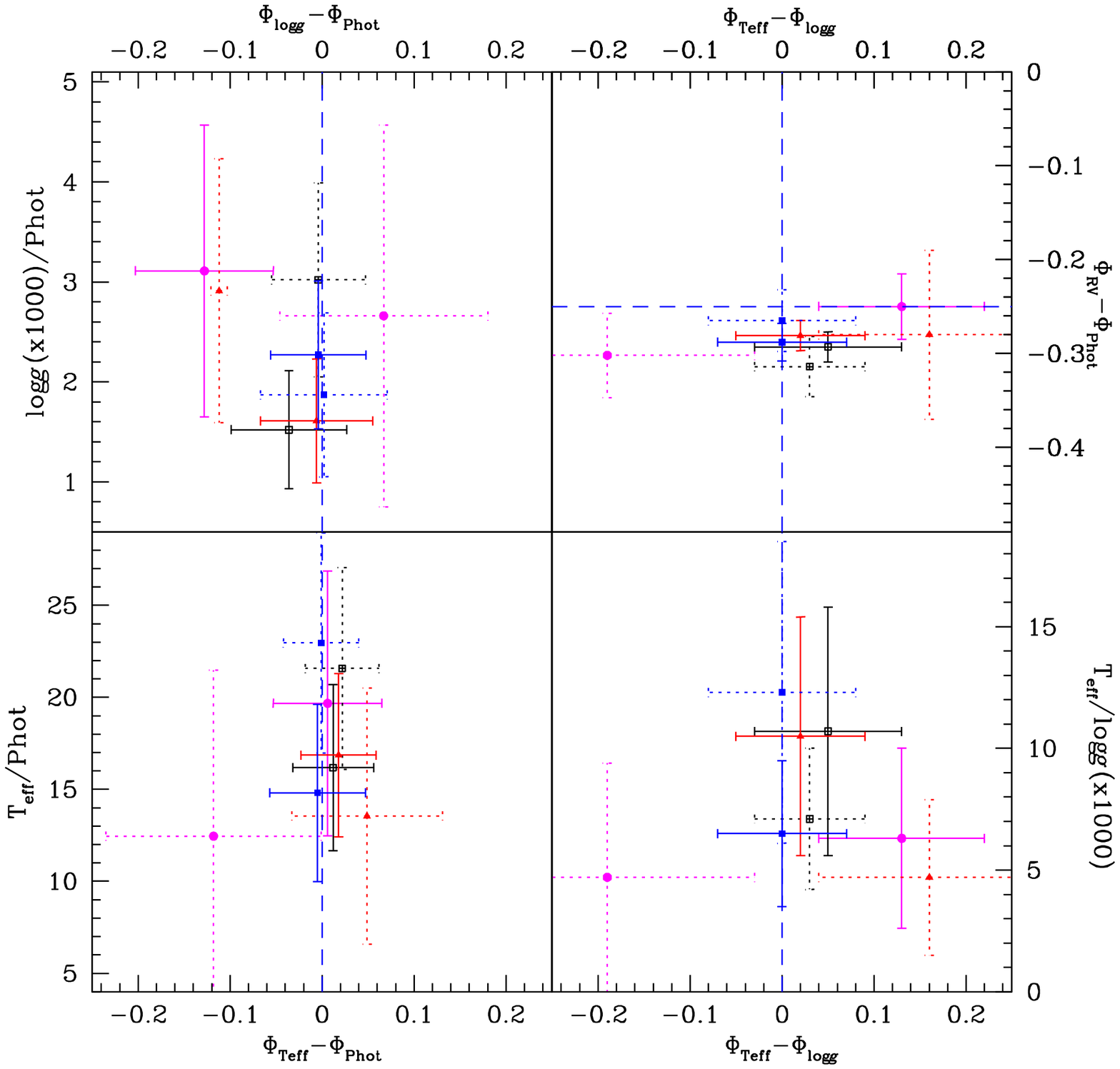}
\caption{Same as Fig.~\ref{fig16} comparing photometry and  model-fitted 
quantities.}
\label{fig18}
\end{figure}

Figures~\ref{fig16} through \ref{fig18} show amplitude ratios and
phase differences of the observables using most of
the available combinations. Figure~\ref{fig16} shows those quantities
directly obtained from the data, Fig.~\ref{fig17} excludes photometry,
but includes spectroscopic quantities derived from models, and Fig.\ref{fig18}
compares the photometry with the model quantities. Values for the NOT/BO
data have solid errorbars while those for the KPNO/MDM data have dotted
errorbars.
We note that most of our quantities are
within the $1\sigma$ errors of all other quantities, when the 
largely instrumental EW offset
between data sets is accounted for.
While we do not attempt any
mode identifications here, the similarities exhibited between the amplitude
ratios suggest low degree modes with $\ell\leq 2$, which can behave
similarly at most inclinations. The $\ell =3$ identification
for $f3$ seems unlikely since this frequency behaves very much like the
others. However, there are no published model amplitude ratios by which
to judge our results.

There are some expected relationships in phases that can be used to
deduce the accuracy of the data as well. It is known that in low-degree modes,
RV phases should be separated by one quarter ($\pi /2$) 
from $T_{\rm eff}$, $\log g$, EW and
photometric phases for adiabatic pulsations. This has been
observed for other sdBV stars \citep{to04,till}, and we see the same
thing. It should also be
expected that brightness, $T_{\rm eff}$, and $\log g$ should all be in
phase but EW should be in anti-phase to these observables, and again, this is
what we observe.
Expected phase relations, deduced from atmospheric models are indicated as 
dashed lines in the figures. Our observations scatter tightly around these
lines, indicating that our spectroscopic reductions and model fitting 
are reliable.

\section{Conclusions}
During 2006 we collected simultaneous photometry and time-series
spectroscopy from MDM and Kitt Peak Observatories and Baker Observatory
and the NOT of the pulsating subdwarf B star PG~1219+534. 
In total, we obtained over 5000 spectra and nearly 15000
photometric measurements.
We have
recovered the four frequencies of the discovery paper \citep{koen99b}
in photometric, velocity, and equivalent width variations. 

From four years of photometry, we determine that PG~1219 has four consistent
frequencies with occasional low-amplitude transients. One such transient
occurred during our spectroscopic runs, though we did not recover that
frequency in any of our spectroscopically-determined quantities. The
photometric amplitudes of $f1$, $f2$, and $f4$ remain stable over the 
course of our observations but the amplitude of $f3$ increased by 43\% 
between our MDM and BO observations.
Similarly, the photometric phases were stable during our observations, but
most varied between the runs. That of $f3$ showed the largest change in phase 
of 13\% while $f2$ and $f4$ had changes of 4 and 3\%, respectively. Only
$f1$ had no phase changes, to within the errors.

Our spectroscopic results were similar, in that only $f3$ showed significant
amplitude variations between the runs. It has a very low RV amplitude
of 0.8~km/s in the KPNO data, which grows to an amplitude of 3.0~km/s
in the NOT data.  However, the differences in normalized EW amplitudes 
are within the errors for \emph{all} frequencies.

By folding the
spectra over the pulsation period, we were able to fit atmospheric
models to the higher S/N binned spectra to extract $T_{\rm eff}$ and $\log g$
variations for all four frequencies. 
Again all pulsation amplitudes are consistent between the KPNO and
NOT data, except for $f3$ in $T_{\rm eff}$ which more than doubles.
As such, $f3$ has significant amplitude increases of 43, 275, and 124\%
between the KPNO/MDM and NOT/BO observations for photometry, RV, and
$T_{\rm eff}$, respectively. While there is a 33\% increase in EW amplitude
between the runs for $f3$, it is just within the $1\sigma$ errors.

We examined amplitude ratios and phase differences for various
measurables for each frequency. We will interpret these more fully in
a subsequent paper, but observationally, we can conclude that the
ratios and differences do not differ greatly  between frequencies. The
simplest interpretation of this result is that they are all low-degree
($\ell\leq 2$) modes. We also compared phase differences to those
expected from adiabatic atmospheric models. Our measured differences
match those expected, except for small departures which are most likely
caused by nonadiabatic effects.

A large arsenal of  quantities can now be applied to
identify the modes of pulsation. A future paper will match our 
measured quantities
to those from perturbed synthetic pulsation spectra to constrain the
modes of each frequency. This work is also the first successful application
of time-series spectroscopy to a ``normal'' amplitude sdBV star. 
All previous detections were of
stars with unusually high photometric amplitudes \citep{simon0,simon2,simon3,simon,woolf,to04,to06,till} and so our work indicates that such studies can provide
useful measurements for the majority of sdBV stars.

\begin{acknowledgements}
We would like to thank Simon Jeffery for some of the atmospheric models
and fitting routines used in our
analysis and  the NOT, MDM, and KPNO TACs for time allocations.
MDR was
supported by an American Astronomical Society Small Research Grant and
the National Science Foundation Grant AST007480. 
Any opinions, findings, and conclusions or
recommendations expressed in this material are those of the
author(s) and do not necessarily reflect the views of the American 
Astronomical Society or the National
Science Foundation. MDR would also like to thank Conny Aerts and a HELAS 
travel grant which allowed us to congregate in Leuven to work on atmospheric
models.
JRE, SLH, and RLVW were supported by the Missouri Space Grant Consortium
and an REU Supplement grant from the National Science Foundation.
R.{\O}. is supported by the Research Council of the University of Leuven and
the FP6 Coordination Action HELAS of the EU.

Some of the data presented here have been taken using {\sc alfosc}, which 
is owned by the Instituo de Astrofisica de Andalucia (IAA) and operated
at the Nordic Optical Telescope under agreement between IAA and the NBIfAFG
of the Astronomical Observatory of Copenhagen.

Based on observations made with the Nordic Optical Telescope, operated on
the island of La Palma jointly by Denmark, Finland, Iceland, Norway, and
Sweden, in the Spanish Observatorio del Roque de los Muchachos of the 
Instituto de Astrofisica de Canarias.
\end{acknowledgements}

\end{document}